\documentclass{article}
\usepackage[english]{babel}
\usepackage{amsmath,amssymb,tabularx,bm,array,booktabs,authblk}
\usepackage{xcolor,slashed,epsfig,cite,float}
\usepackage{verbatim,slashed,ulem,geometry,setspace,algorithm}
\usepackage{algorithmic,mathrsfs,braket,multirow,dcolumn,subfigure}
\usepackage{graphicx}
\usepackage{threeparttable}
\usepackage[colorlinks=true,citecolor=blue,urlcolor=cyan,filecolor=magenta,linkcolor=red,frenchlinks=true,bookmarks]{hyperref}
\newcommand{\nobracket}{}

\newcommand{\email}[1]{\thanks{\href{mailto:#1}{#1}}}

\begin{document}
\title{Transversity Generalized Parton Distributions of $\Delta$ \\ with the Diquark Spectator Model}

\author[a]{Dongyan Fu\email{fudongyan@impcas.ac.cn}}
\author[b,c]{Yubing Dong\email{dongyb@ihep.ac.cn}}
\author[a,d,e]{S. Kumano\email{kumanos@impcas.ac.cn}}

\affil[a]{Southern Center for Nuclear-Science Theory,
\authorcr
Institute of Modern Physics, Chinese Academy of Sciences, Huizhou 516000, China}
\affil[b]{Institute of High Energy Physics, Chinese Academy of Sciences, 
Beijing 100049, China}
\affil[c]{School of Physical Sciences, University of Chinese Academy of Sciences, Beijing 101408, China}
\affil[d]{Quark Matter Research Center, Institute of Modern Physics, Chinese Academy of Sciences,
Lanzhou, 730000, China}
\affil[e]
{KEK Theory Center, Institute of Particle and Nuclear Studies, KEK,
Oho 1-1, Tsukuba, 305-0801, Japan}

\maketitle

\begin{abstract}
We show quark transversity generalized parton pistributions (GPDs) of $\Delta^+$ isobar by using 
the diquark spectator model for the first time. 
First, this model is tested by electric charge, magnetic-dipole and axial charge form factors, and it is used for calculating
the transversity GPDs $H^{qT}_{1,3,5,7}$ of $\Delta^+$.
The quark transversity distribution $h_1$ is then obtained 
from the transversity GPDs in the forward limit. 
Then, helicity-flip amplitudes are shown numerically
by using relations between the helicity amplitudes and the GPDs.
Finally, by taking first moments of the GPDs, tensor form factors 
are obtained and we predict the tensor charge.
Experimentally, $N$-$\Delta$ transition GPDs are investigated 
in deeply virtual Compton scattering and virtual meson-production
processes, and generalized distribution amplitudes, which correspond 
to the $s$-channel GPDs, could be investigated by 
the two-photon processes $\gamma^* \gamma \to \Delta \bar\Delta$
at the electron-positron colliders.
Therefore, the spin-3/2 $\Delta$ GPDs could become interesting quantities
experimentally in future.
\end{abstract}

\maketitle

\section{\uppercase{Introduction}}

Generalized parton distributions (GPDs) have become important 
functions to describe the hadron structure since 
they were proposed~\cite{Ji:1998pc,Ji:1996nm,Radyushkin:1997ki,
Muller:1994ses}. The GPDs contain abundant information about 
quark and gluon distributions inside the hadron, besides the usual 
parton distributions. It is known that the GPDs have three variables: 
the parton longitudinal momentum fraction $x$, the transferred 
momentum square $t$, and the skewness $\xi$, which is defined by 
the longitudinal transferred momentum.
On the other hand, parton-hadron helicity amplitudes, which 
can be expressed by the GPDs, constraint that there are $2(2 J+1)^2$ 
independent GPDs for a quark or a gluon of a spin-$J$ hadron 
at the leading twist. 
The half of the GPDs are helicity nonflip GPDs which
are unpolarized and longitudinal polarized GPDs,
and another half are helicity flip GPDs, $i$.$e$. transversity GPDs.
For a spin-$\frac{3}{2}$ hadron, there are 16 independent 
transversity GPDs, and these GPDs are chiral-odd 
since the non-local operator flips the quark chirality in contrast 
to the chiral-even unpolarized and longitudinal polarized GPDs.

Similar to the deep inelastic scattering process that can be used
to measure the parton distributions, off-forward Compton scattering reaction 
can be utilized to measure the GPDs, like the well-known deeply 
virtual Compton scattering (DVCS) and deeply virtual meson 
production (DVMP) processes. The DVCS reveals the correlations between 
the momentum and spatial degrees of freedom of the partons and 
these correlations contain a wealth of information on the 
structure of hadrons. In the recent years, the proton DVCS has 
been measured by several accelerator facilities such as Jefferson Lab~\cite{CLAS:2001wjj,CLAS:2006krx,CLAS:2007clm,
JeffersonLabHallA:2007jdm,CLAS:2008ahu,CLAS:2014qtk,CLAS:2015bqi,
JeffersonLabHallA:2015dwe,CLAS:2015uuo,CLAS:2018ddh}, 
and DESY~\cite{HERMES:2001bob,ZEUS:2003pwh,H1:2005gdw,
HERMES:2006pre,ZEUS:2008hcd,HERMES:2012gbh}.
In addition, there are future GPD projects at 
EicC~\cite{Anderle:2021wcy} and EIC~\cite{AbdulKhalek:2021gbh}, 
while Japan Proton Accelerator Research Complex (J-PARC)
~\cite{Kumano:2009he,Qiu:2022bpq} and 
Fermilab~\cite{Kumano:2022ptt,Chen:2024adg} also have 
possibilities to measure GPDs.
There have been many theoretical studies about the GPDs, especially 
the spin-0 pion~\cite{Praszalowicz:2002ct,Ji:2006ea,
Frederico:2009fk,Fanelli:2016aqc,Raya:2021zrz}, the 
spin-$\frac{1}{2}$ proton~\cite{Polyakov:2002yz,Mineo:2005vs,
Chakrabarti:2013gra,Diehl:2013xca,Mondal:2015uha,
Chakrabarti:2015ama} and the spin-1 
$\rho$~\cite{Sun:2017gtz,Sun:2018ldr} and 
deuteron~\cite{Berger:2001zb,Dong:2013rk,Cosyn:2018rdm}. 

On the other hand, the $s$-$t$ crossed process of the DVCS
$\gamma^* h \to \gamma h$ with the photon $\gamma$ and
a hadron $h$ is the two-photon process $\gamma^* \gamma \to h \bar h$.
Although the hadron $h$ should be generally a stable one in the DVCS,
the $h$ could be an unstable hadron in the two-photon process.
It enables us to investigate internal structure of 
unstable hadrons with higher spins such as the spin-1 $\rho$
and spin-3/2 $\Delta$. In the two-photon processes \cite{PhysRevD.93.032003},
generalized distribution amplitudes (GDAs) are investigated 
\cite{PhysRevD.97.014020}, and they could be considered as
$s$-channel GPDs. In fact, the GDAs of the pion were investigated
and its gravitational form factors were extracted from 
experimental data \cite{PhysRevD.97.014020}.
The unstable hadron GPDs can be also investigated
in the form of transition GPDs \cite{Diehl2025}.
In fact, there are already experimental data on
the $N \to \Delta$ transition at
Jefferson Lab and it will be investigated by the future EICs.
Under the development of these experimental techniques, 
it is increasing important to study structure functions
of higher-spin hadrons including the spin-3/2 $\Delta$.
The GDAs ($s$-channel GPDs) of $\Delta$ could be measured 
by $\gamma^* \gamma \to \Delta \bar \Delta$ and
the transition GPDs were already measured for the $N \to \Delta$.
In this work, we study the transversity GPDs of 
the spin-$\frac{3}{2}$ particles numerically.

In our previous work~\cite{Fu:2022bpf,Fu:2024kfx,Fu:2023dea}, we 
have given the decomposition of the spin-$\frac{3}{2}$ twist-2 
GPDs including unpolarized, longitudinal polarized, and 
transversity polarized GPDs, and the numerical calculations of the 
quark unpolarized and longitudinal polarized parts have been 
performed. In the preset work, we show the quark transversity GPDs
numerically by using the diquark spectator model.
In the forward limit, the transversity GPDs give 
the transversity distribution function 
$h_1(x)$, which is explained as the number density of a parton 
with the longitudinal momentum fraction $x$ and polarization 
parallel to that of the hadron with transverse polarized minus 
the number density with the antiparallel polarization.
The transversity distribution function $h_1(x)$ has been 
explored by many studies~\cite{Hoodbhoy:1998vm,Cosyn:2018rdm,Jaffe:1989xy,Diehl:2001pm,Ralston:1979ys,Artru:1989zv,Jaffe:1991kp,Jaffe:1991ra,Cortes:1991ja,Ji:1992ev,Barone:2001sp}.
Moreover, the first moments of the transversity GPDs 
are tensor form factors of $\Delta$,
and numerical results of these form factors are also 
given in this paper.

In the preceding paper \cite{Fu:2024kfx}, we showed the general formalism
what kind of the transversity GPDs exist for spin-3/2 hadrons.
The purpose of this paper is to show the quark transversity GPDs of $\Delta$
numerically by uing the diquark spectator model.
This work is intended to show the magnitude and functional forms
of $x$ and $t$ of the transversity GPDs for possible future
theoretical and experimental studies.
In this paper, we make a qualitative estimate for the transversity GPDs 
of spin-$\frac{3}{2}$ particles, taking $\Delta^+$ as an example. 
In Sec.~\ref{section2}, the general formalism is given
for the transversity GPDs and corresponding tensor form factors.
Then, the diquark spectator model is introduced.
In Sec.~\ref{section3}, numerical results are shown for the transversity GPDs, the parton distribution functions (PDFs) of $\Delta^+$,
helicity-flip amplitudes, and tensor form factos.
Finally, the summary is given in Sec.~\ref{section4}.

\section{\uppercase{transversity GPDs for spin-3/2 hadrons and the diquark spectator model}}\label{section2}

\subsection{Transversity GPDs for spin-3/2 hadrons}

\begin{figure}[ht]
    \centering
    \includegraphics[height=2cm]{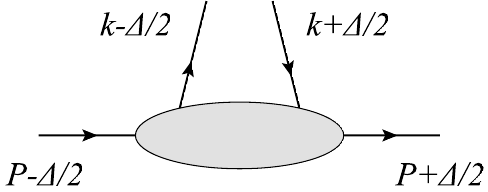}
    \caption{\small{Diagram describing the quark GPDs.}}
    \label{quarkgpdfigure}
\end{figure}

In this work, the leading twist quark transversity GPDs of the spin-3/2 hadron, $\Delta$, 
are calculated for the first time by using the diquark spectator model. 
We take the $d$ quark GPDs in $\Delta^+$ as an example, and we can obtain 
the numerical results of other quarks
by counting the corresponding quark number. 
The basic formalism is shown in this section for 
the transversity GPDs of the spin-3/2 hadrons \cite{Fu:2024kfx}.
The transversity GPDs of the hadrons are defined through the matrix element 
of the non-local quark operator with spin-flip:
\begin{equation}\label{quarktransversitygpds}
	T^{q i}_{\lambda' \lambda}=\frac{1}{2} \int \frac{\text{d} z^-}{2 \pi}
	e^{i x \left( P \cdot z \right)} \left.
	\left\langle p', \lambda' \left| \bar{\psi} (-\frac{1}{2}z) i \sigma^{n i} \psi
	(\frac{1}{2}z) \right|p,
	\lambda \right\rangle \right| _{z^+=0, \bm{z}_\perp=0},
\end{equation}
The $p$ ($p'$) and $\lambda$ ($\lambda'$) respectively denote the momentum and helicity
of the initial (final) state, and $\lambda, \, \lambda' = \pm \frac{3}{2},\,\pm \frac{1}{2}$ 
for spin-3/2 hadrons. 
The light-cone coordinate is employed and any four-vector $v$ can be rewritten 
as $v=(v^+,v^-,\bm{v}_\perp)$, where $v^\pm = v^0 \pm v^3$ and $\bm{v}_\perp=(v^1, v^2)$. 
The scalar product of any two four-vectors then is 
$u \cdot v=\frac{1}{2} u^+ v^- +\frac{1}{2} u^- v^+ -\bm{u}_\perp \cdot \bm{v}_\perp$.
Moreover, the light-cone vector $n=(0,2,\bm{0}_\perp)$ is needed and $n^2=0$.
In addition, we use the same kinematical variables 
with our previous studies~\cite{Fu:2024kfx},
\begin{equation}
	P=\frac{p'+p}{2}, \quad \Delta=p'-p, \quad t=\Delta^2, \quad \xi =
	- \frac{\Delta^+}{2 P^+} \,(|\xi| \leq 1),
	\quad x=\frac{k^+}{P^+}\,(-1 \leq x \leq 1),
\end{equation}
where $k-\Delta/2$ ($k+\Delta/2$) is the initial (final) parton momentum as displayed 
in Fig.~\ref{quarkgpdfigure}.
Here, we use the same symbol $\Delta$ to represents the momentum transfer and 
the hadron name. 
The following conventions, $a^{[ \mu \nobracket} b^{\nobracket \nu ]}=a^\mu b^\nu - a^\nu b^\mu$,
$a^{\{ \mu \nobracket} b^{\nobracket \nu \}}=a^\mu b^\nu + a^\nu b^\mu$,
$\sigma^{n i}=\sigma^{\rho i} n_{\rho}$,
$\sigma^{\mu\nu} = (i/2) [\gamma^\mu,\gamma^\nu]$,
$\epsilon^{n i \rho \delta}=\epsilon^{\mu i \rho \delta} n_\mu$, and $\epsilon_{0123}=1$, are used.
Moreover, $i$ in the definition~\eqref{quarktransversitygpds} is the transverse index, $i=1,\, 2$.

The quark transversity GPDs are defined by the matrix elements of
the transverse non-local quark-quark correlator in Eq.~\eqref{quarktransversitygpds} as
\begin{equation}\label{quarkh}
    T^{q i}_{\lambda' \lambda} = -\bar{u}_{\alpha'} (p',\lambda')
    \mathcal{H}^{q T, i, \alpha' \alpha}(x,\xi,t) u_\alpha(p,\lambda),
\end{equation}
where $u_\alpha(p,\lambda)$ is the spin-3/2 field Rarita-Schwinger spinor, shown
in Appendix A of Ref.~\cite{Fu:2024kfx}, normalized to
$\bar{u}_{\alpha} (p,\lambda') u^\alpha(p,\lambda)=-2 M \delta_{\lambda' \lambda}$.
Hermiticity, parity invariance, and time-reversal invariance imply
the 16 independent transversity GPDs decomposed from the tensor function 
$\mathcal{H}^{q T, i, \alpha' \alpha}(x,\xi,t)$ as
\begin{equation}\label{quarkdecompositions}
	\begin{split}
		\mathcal{H}^{q T, i, \alpha' \alpha}= & H^{q T}_1 \frac{i \sigma^{n i}}
		{\left( P \cdot n \right)} g^{\alpha' \alpha}
		+ H^{q T}_2 \frac{n^{[ \alpha' \nobracket} g^{\nobracket \alpha ] i}}
		{\left( P \cdot n \right)}
		+ H^{q T}_3 \frac{\left( \slashed{n} P^i - P \cdot n \, \gamma^i \right)}
		{M \left( P \cdot n \right)} g^{\alpha' \alpha}
		+ H^{q T}_4 \frac{\left( \slashed{n} P^i - P \cdot n \, \gamma^i \right)}
		{M^3 \left( P \cdot n \right)} P^{\alpha'} P^\alpha \\
		& + H^{q T}_5 \frac{\left( \slashed{n} \Delta^i - \Delta \cdot n \,
		\gamma^i \right)}{M \left( P \cdot n \right)} g^{\alpha' \alpha}
		+ H^{q T}_6 \frac{\left( \slashed{n} \Delta^i - \Delta \cdot n \, \gamma^i \right)}{M^3
		\left( P \cdot n \right)} P^{\alpha'} P^\alpha \\
		& + H^{q T}_7 \frac{\left(\Delta^i + 2 \xi P^i \right)}{M^2}g^{\alpha' \alpha}
		+ H^{q T}_8 \frac{\left(\Delta^i + 2 \xi P^i \right)}{M^4}P^{\alpha'} P^\alpha\\
		& + H^{q T}_{9} \frac{\left( \Delta \cdot n \, n^{\{ \alpha' \nobracket}
		g^{\nobracket \alpha \} i} -2 n^{\alpha'}
		n^\alpha \Delta^i \right)}{\left( P \cdot n \right)^2}
		+ H^{q T}_{10} \frac{\left( P \cdot n \, n^{\{ \alpha' \nobracket}
		g^{\nobracket \alpha \} i}
		-2 n^{\alpha'} n^\alpha P^i \right)}{\left( P \cdot n \right)^2}\\
		& + H^{q T}_{11} \frac{\left( \Delta \cdot n \, P^{[ \alpha' \nobracket}
		g^{\nobracket \alpha ] i}- P^{[ \alpha' \nobracket} n^{\nobracket \alpha ]}
		\Delta^i \right)}{M^2 \left( P \cdot n \right)}
		+ H^{q T}_{12} \frac{\left( P \cdot n \, P^{[ \alpha' \nobracket} g^{\nobracket \alpha ] i}
		- P^{[ \alpha' \nobracket} n^{\nobracket \alpha ]} P^i \right)}
		{M^2 \left( P \cdot n \right)}\\
		& + H^{q T}_{13} \frac{M \slashed{n} \left( \Delta \cdot n n^{\{ \alpha' \nobracket}
		g^{\nobracket \alpha \} i} - 2 n^{\alpha'}
		n^\alpha \Delta^i \right)}{\left( P \cdot n \right)^3}
		+ H^{q T}_{14} \frac{M \slashed{n} \left( P \cdot n n^{\{ \alpha' \nobracket}
		g^{\nobracket \alpha \} i} - 2 n^{\alpha'} n^\alpha P^i \right)}
		{\left( P \cdot n \right)^3}\\
		& + H^{q T}_{15} \frac{\slashed{n} \left( \Delta \cdot n \, P^{[ \alpha' \nobracket}
		g^{\nobracket \alpha ] i}- P^{[ \alpha' \nobracket} n^{\nobracket \alpha ]}
		\Delta^i \right)}{M \left( P \cdot n \right)^2}
		+ H^{q T}_{16} \frac{\slashed{n} \left( P \cdot n \, P^{[ \alpha' \nobracket}
		g^{\nobracket \alpha ] i}- P^{[ \alpha' \nobracket} n^{\nobracket \alpha ]}
        P^i \right)}{M \left( P \cdot n \right)^2},
	\end{split}
\end{equation}
where the variables $x$, $\xi$, $t$ in the quark transversity GPDs $H^{q T}_i$ are omitted, 
and $M$ is the mass of the spin-3/2 hadron.
Here, $H^{q T}_{3,4,10,11,14,15}$ are $\xi$-odd and others are $\xi$-even with respect
to the skewness $\xi$ as follows
\begin{equation}\label{odd-even}
	\begin{split}
		H^{q T}_i(x,\xi,t) & =H^{q T}_i(x,-\xi,t) \quad \text{with}
		\quad i=1,2,5 \sim 9,12,13,16,\\
		H^{q T}_j(x,\xi,t) & =-H^{q T}_j(x,-\xi,t) \quad \text{with} \quad j=3,4,10,11,14,15.\\
	\end{split}
\end{equation}
Then, all the $\xi$-odd transversity GPDs vanish in the 
limit $\xi \rightarrow 0$. In our calculation with the diquark spectator model, only the negative $\xi$ will be 
considered. Then, the positive part can be obtained by the parity of GPDs about $\xi$.
In the forward limit, the transversity GPDs change to transversity distribution functions $h_1(x)$,
\begin{equation}\label{quarktransversitydistribution}
    2 \, [H^{q T}_1(x,0,0)-H^{q T}_2(x,0,0)]=h_1(x),
\end{equation}
where the number density is from two components, spin-1/2 and spin-1, due to the fact that the
Rarita-Schwinger spinor is composed  by the fields of spin-1/2 and spin-1.

The non-local tensor quark-quark operator gives one local tensor current 
using the sum rule according to the Mellin moment,
\begin{equation}
	\begin{aligned}
		& (P \cdot n)^{a + 1} \int \text{d} x \, x^a \int \frac{\text{d} z^-}{2 \pi} 
		e^{i x P^+ z^-} \left[\, \overline{\psi}
		\left( - z/2 \right) i \sigma^{n \nu}
		\psi \left( z/2 \right) \right]
		\Bigg |_{z^+ = 0, \mathbf{z} 
		= 0}\\
		= & \left( i \frac{\text{d}}{\text{d} z^-} \right)^a \left[\, \overline{\psi}
		\left( - z/2 \right)i \sigma^{n \nu}
		\psi \left( z/2
		\right) \right] \Bigg |_{z = 0} = \overline{\psi} (0) 
		i \sigma^{n \nu} (i\overleftrightarrow{\partial}^+)^a \psi (0) .
	\end{aligned} \label{Mellin}
	\end{equation}
Then, the corresponding tensor form factors can be defined using the local tensor current, 
which is connected to the $1$th Mellin moment ($a=0$) in $x$, as
\begin{equation}
	\begin{split}\label{tensorcurrent}
		T^{\mu \nu}&=\left\langle p', \lambda' \left| \bar{\psi} (0)
		i \sigma^{\mu \nu} \psi (0) \right|p,\lambda \right\rangle \\
		&=-2\,\bar{u}_{\alpha'} (p',\lambda') \mathcal{F}_q^{\mu \nu,\alpha' \alpha} u_\alpha(p,\lambda),
	\end{split}
\end{equation}
where the tensor form factors are $\xi$-independent. 
In the previous work~\cite{Fu:2024kfx}, the explicit decomposition of the matrix element 
of the tensor current~\eqref{tensorcurrent} is given as
\begin{equation}\label{tensorformfactors}
	\begin{split}
		\mathcal{F}_q^{\mu \nu,\alpha' \alpha}= & g^{\alpha' \alpha}\left(G^{q T}_1(t)
		i \sigma^{\mu \nu} + G^{q T}_5(t) \frac{\gamma^{[ \mu \nobracket}
		\Delta^{\nobracket \nu ]}}{M}
		+ G^{q T}_7(t) \frac{P^{[ \mu \nobracket} \Delta^{\nobracket \nu ]}}{M^2} \right) \\
		& + \frac{P^{\alpha'} P^\alpha}{M^2}\left( G^{q T}_6(t)
		\frac{\gamma^{[ \mu \nobracket} \Delta^{\nobracket \nu ]}}{M} + G^{q T}_8(t)
		\frac{P^{[ \mu \nobracket} \Delta^{\nobracket \nu ]}}{M^2} \right)
		+ G^{q T}_{2}(t) g^{\mu [ \alpha' \nobracket} g^{\nobracket \alpha ] \nu}
		+ G^{q T}_{12}(t) \frac{P^{[\alpha' \nobracket} g^{\nobracket \alpha ]
		[\nu \nobracket} P^{\nobracket \mu ]}}{M^2}.
	\end{split}
\end{equation}
One can then obtain the sum rules connected the tensor FFs $G^{q T}_i(t)$ 
with the transversity GPDs as
\begin{equation}\label{quarksumrules}
\begin{split}
    \int ^1_{-1} d x \, H^{q T}_i(x,\xi,t)&=G^{q T}_i (t) \quad \text{with}
    \quad i=1,2,5 \sim 8,12,\\
    \int ^1_{-1} d x \, H^{q T}_j(x,\xi,t)&=0 \quad \text{with}
    \quad j=3,4,9,10,11,13 \sim 16.
\end{split}
\end{equation}
Moreover, the combination $G_{1}^{q T}(0) - G_{2}^{q T}(0)$ describes 
the quark tensor charge carried by the corresponding quark.

\subsection{Diquark spectator model}\label{section-model}

In the present work, we consider the $\Delta^+$ as an attempt 
to characterize the multidimensional structure 
of the spin-3/2 particles.
In the picture of the quark model, the $\Delta^+$ isobar is composed 
of three light quarks, two $u$ quarks and one $d$ quark.
The quantum numbers of $\Delta^+$ are $I \, (J^P)=3/2 \, (3/2^+)$ 
and it requires that both the isospin and spin of each pair of quarks be 1. 
Therefore, it is convenient to regard two quarks in $\Delta^+$ as a whole, 
$i$.$e$. diquark.
We treat both quark and diquark 
as elementary particles, respectively.
However, the calculations of the diquark GPDs are difficult 
because of more complicated integrals, so we calculate 
the GPDs for each flavor quark by using the diquark spectator model 
instead of the quark-diquark approach.

\begin{figure}[htbp]
	\centering
	\includegraphics[height=3cm]{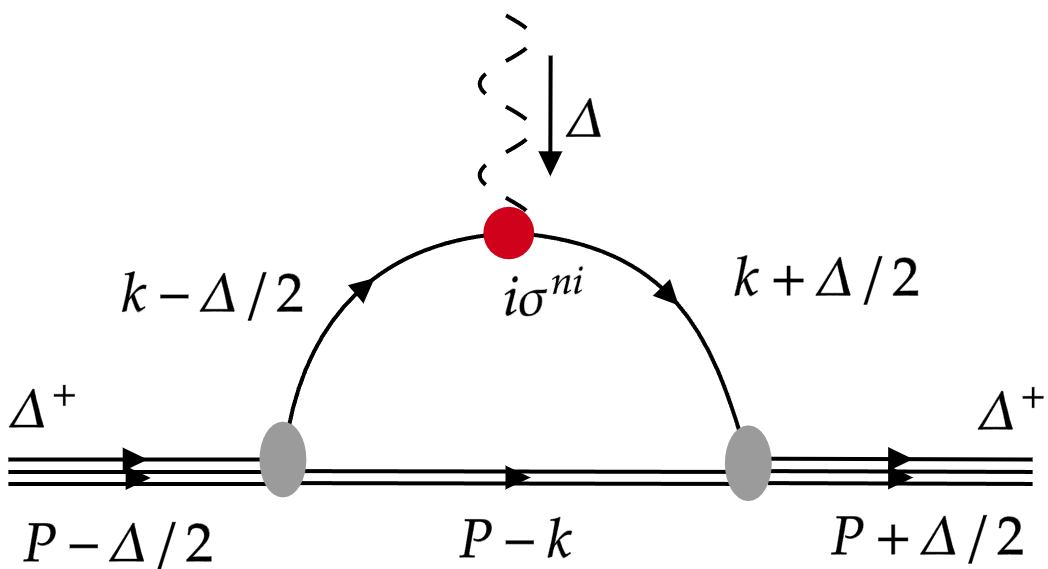}
	\caption{\small{Feynman diagram for the $\Delta^+$ GPDs 
	using the diquark spectator approach, and the single (double) line stands
	for the quark (diquark).}}
	\label{feynmandiagram}
\end{figure}

The Breit frame, $\Delta^+ = - \Delta^-$, is employed for convenience in this work, 
where the initial and final momenta are
\begin{equation}
	p=(P^0-\frac{\Delta_z}{2}, P^0+\frac{\Delta_z}{2}, -\frac{\bm{\Delta}_\perp}{2}),\quad
	p'=(P^0+\frac{\Delta_z}{2}, P^0-\frac{\Delta_z}{2}, \frac{\bm{\Delta}_\perp}{2}).
\end{equation}
Figure \ref{feynmandiagram} displays the Feynman diagram of the diquark spectator approach.
The quark transversity GPDs defined in Eq.~\eqref{quarkh} are calculated 
according to the Feynman diagram as 
\begin{equation}\label{unint1}
	\mathcal{H}^{\alpha' \alpha}
		= -\frac{i}{2} c_1^2 \int \frac{\text{d}^2 k_\perp \text{d} k^+ \text{d} k^-}{(2 \pi)^4}
		\frac{\delta(k\cdot n - x P \cdot n)}{\mathfrak{D}}
		\Gamma^{\alpha' \beta'} \left(\slashed{k}+\frac{\slashed{\Delta}}{2}+m_q \right)
		g_{\beta' \beta} \slashed{n} \left(\slashed{k}-\frac{\slashed{\Delta}}{2}
		+m_q \right) \Gamma^{\beta \alpha},
\end{equation}
where
\begin{equation}\label{OmegaD}
    \begin{split}
    \mathfrak{D}=& \left[\left(k+ \frac{\Delta}{2} \right)^2 - m_q^2+i \epsilon\right]
    \left[\left(k- \frac{\Delta}{2} \right)^2 - m_q^2+i \epsilon\right]
	\left[\left(k-P\right)^2-m_D^2+i \epsilon \right] \left[\left( k-P \right) ^2 -m_R^2+i \epsilon \right]^2\\
    & \times
    \left[\left( k+\frac{\Delta}{2} \right) ^2 -m_R^2+i \epsilon\right]
	\left[\left( k-\frac{\Delta}{2} \right) ^2 -m_R^2+i \epsilon\right],
    \end{split}
\end{equation}
and the effective form of the vertex function is \cite{Scadron:1968zz}
\begin{equation}\label{vertexfunction}
		\Gamma^{\alpha \beta} = g^{\alpha \beta} + c_2 \gamma^\beta \Lambda^\alpha
		         + c_3 \Lambda^\beta \Lambda^\alpha,
\end{equation}
where $\Lambda$ is the relative momentum between the spin-1/2 and spin-1 partons. 
In Eq.~\eqref{unint1}, we also employ the same scalar function of the loop momentum, 
$\Xi (p_1,p_2,m_R)=c_1/(p_1^2-m_R^2+i \epsilon)(p_2^2-m_R^2+i \epsilon)$ 
where $p_1$ and $p_2$ respectively represent the momentum of the spin-1/2 
and spin-1 partons, with Ref.~\cite{Fu:2023dea} to perform the regularization. Moreover, we adopt the same simplified diquark propagator $g_{\beta' \beta}$ as used in Ref.~\cite{Fu:2023dea}.
The calculation details are shown in the previous work~\cite{Fu:2022rkn,Fu:2023dea}.

In Eqs.\,(\ref{OmegaD},\ref{vertexfunction}), there are 
model parameters $m_R$ and $c_{1,2,3}$.
In this work, the same parameter $m_R$ in Ref.~\cite{Fu:2023dea} is employed. 
The parameter of $c_1$ is determined by the electric charge.
The parameters $c_{2,3}$ could affect the higher multipole terms
and they are taken zero in this work.
Under this approximation, the vertex becomes $g^{\alpha' \alpha}$,
which implies that the calculation of the spin-3/2 hadron 
will regress to the spin-1/2 condition and only four transversity GPDs 
corresponding to the spin-1/2 will be obtained.
However, the spin-1/2 part is the leading term and 
the following numerical results verify that the spin-1/2 part 
can give the reliable leading results.
Therefore, we think that the transversity PDF and 
the tensor form factors from this simplification 
are also reliable. 
To extract GPDs from the $\mathcal{H}^{q T, i,\alpha' \alpha}$, 
one needs some identities and on-shell identities like Schouten
and Gordon identities which are listed in the Appendixes 
of Refs.~\cite{Fu:2022bpf,Fu:2023dea,Fu:2024kfx}.
One can find the calculation details in our previous works~\cite{Fu:2023dea}.

\section{\uppercase{numerical results}}\label{section3}

\begin{figure}[h]
	\centering
	\includegraphics[width=0.32 \textwidth]{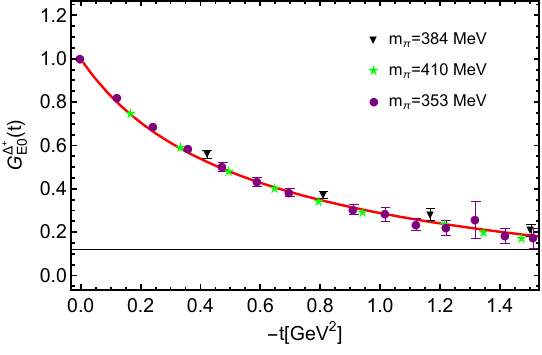}
	\includegraphics[width=0.32 \textwidth]{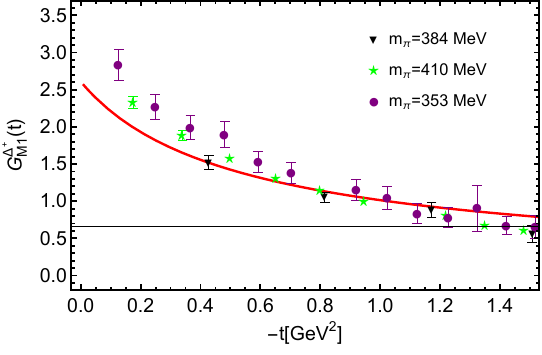}
	\includegraphics[width=0.32 \textwidth]{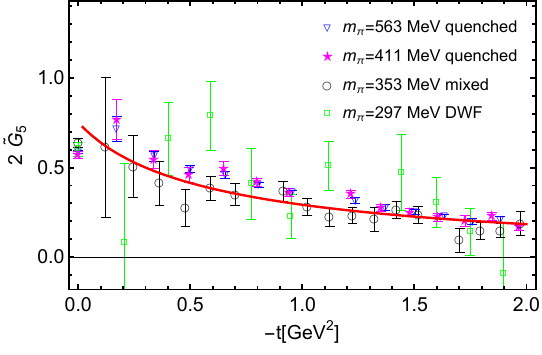}
	\caption{\small{The electric charge, magnetic moment and axial charge form factors of $\Delta^+$ with $c_2=c_3=0$, 
	compared with Ref.~\cite{Alexandrou:2008bn,Alexandrou:2013opa}.}}
	\label{fig-unffs}
\end{figure}

In this section, we present the numerical results about the 
transversity GPDs for $d$ quark of $\Delta^{+}$ as an example. 
We use the same masses, $\Delta$ resonance mass $M$, quark mass 
$m_q$, diquark mass $m_D$ and cutoff mass $m_R$ as in our previous 
papers~\cite{Fu:2022rkn,Fu:2023dea}, which are listed in 
Table~\ref{table-parameters}.
\begin{table}[ht]
	\renewcommand\arraystretch{1.3}
	\caption{\small{Mass parameters used in this work.}} 
	\centering
 \begin{tabular}{ c c c c c c c}
		 \toprule
		 \toprule
		  $M /\text{GeV}$ & $m_q /\text{GeV}$	
		  &  $m_D /\text{GeV}$ & $m_R /\text{GeV}$ 
		  & $c_2 /\text{GeV}^{-1}$	& $c_3 /\text{GeV}^{-2}$\\
		\midrule
		 1.085	& 0.4	& 0.76	& 1.6 & 0	& 0 \\
		\bottomrule
		\bottomrule
	 \end{tabular}
	\label{table-parameters}
\end{table}
In Ref.~\cite{Fu:2022rkn}, we indicated that $c_{2,3}$ 
in Eq.\,(\ref{vertexfunction}) mainly 
contribute to the high order multipole terms, like electric 
quadrupole and magnetic octupole form factors,
and they are chosen as $c_2=c_3=0$ in this work.
To verify that this simplification is reasonable,
unpolarized and longitudinal polarized form factors are shown. 
The electric charge, magnetic moment and axial charge form factors of $\Delta^+$ are shown in Fig.~\ref{fig-unffs}.
From Fig.~\ref{fig-unffs}, we see that the electric charge and axial charge 
form factors are consistent with lattice QCD calculations,
and the magnetic moment form factor is acceptable. 
These results illustrate that the spin-1/2 part $g^{\alpha \beta}$ 
of the vertex $\Gamma^{\alpha \beta}$ in 
Eq.~\eqref{vertexfunction} can 
provide reliable results in the leading terms.
These results indicate that transversity GPDs and 
tensor form factors corresponding to $h_1$ are reliable in the following subsections.


\subsection{Numerical transversity GPDs and helicity amplitudes}

\begin{figure}[h]
	\centering
	\subfigure[]{\includegraphics[height=4.5cm]{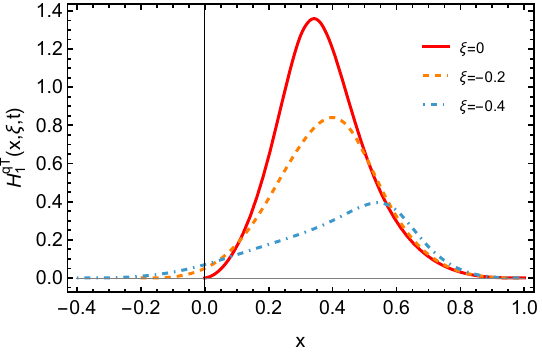}\label{fig-tgpd1}}
    \subfigure[]{\includegraphics[height=4.5cm]{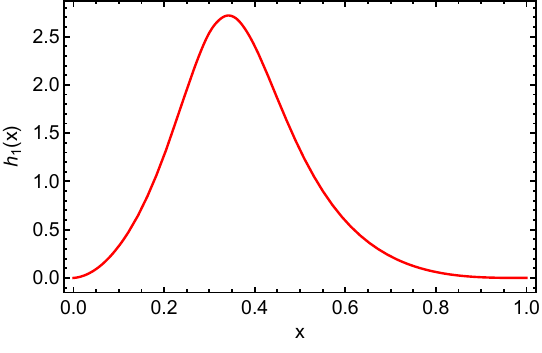}\label{fig-tpdf}}
	\caption{\small{(a): The GPDs $H^{qT}_{1}(x,\xi,-|t|_{\min})$ 
	with different skewness $\xi$ and the corresponding minimal $|t|$. 
	The red solid line, orange dashed line and blue dot dash line respectively 
	represent the GPDs with 
	$\left( \xi, -t \right) = \left( 0, 0 \,\text{GeV}^2 \right)$, 
	$\left( -0.2, 0.2\,\text{GeV}^2 \right) $ 
	and $\left( -0.4, 0.9\,\text{GeV}^2 \right) $. (b): The transversity PDF $h_1(x)$, where $h_1(x) = 2 H_1^{qT}(x,0,0)$.}}
	\label{fig-tpdfsxi}
\end{figure}

As an example, but without loss of generality, the $d$ quark transversity GPDs 
of $\Delta^+$ are given in this section, and
\begin{figure}[H]
	\centering
	\includegraphics[width = 0.2 \textwidth]{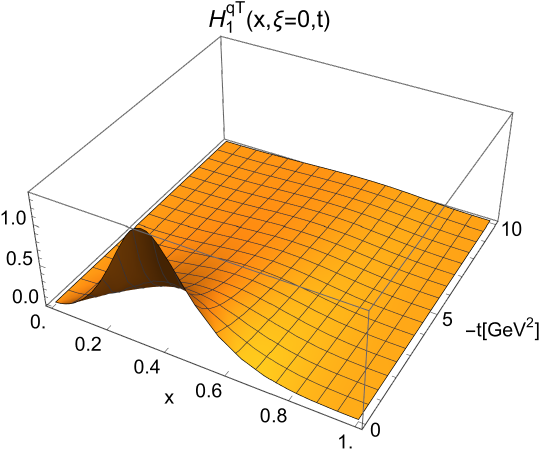}
	\includegraphics[width = 0.2 \textwidth]{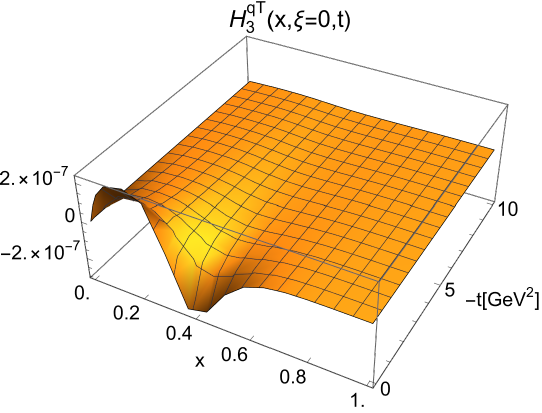}
	\includegraphics[width = 0.2 \textwidth]{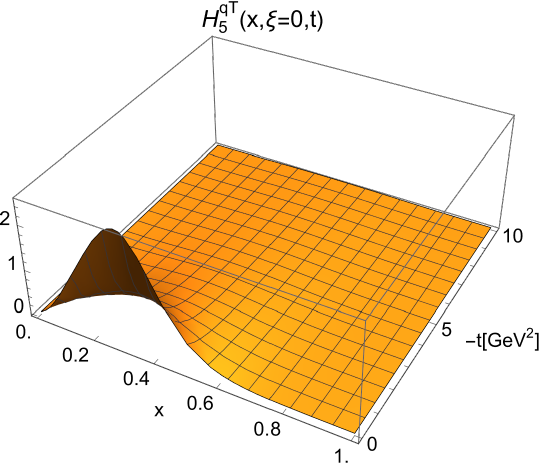}
	\includegraphics[width = 0.2 \textwidth]{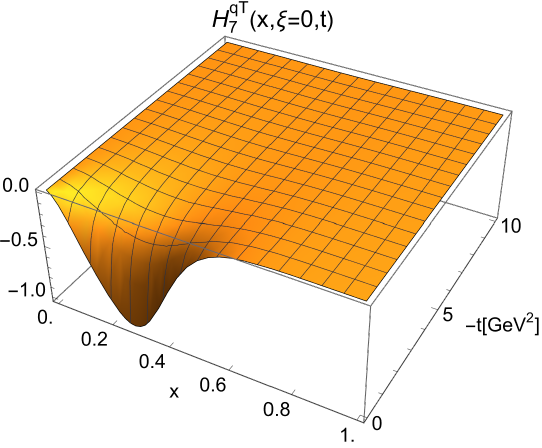}\\
	\includegraphics[width = 0.2 \textwidth]{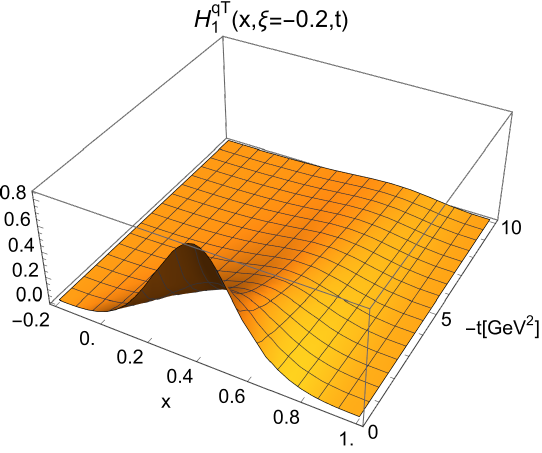}
	\includegraphics[width = 0.2 \textwidth]{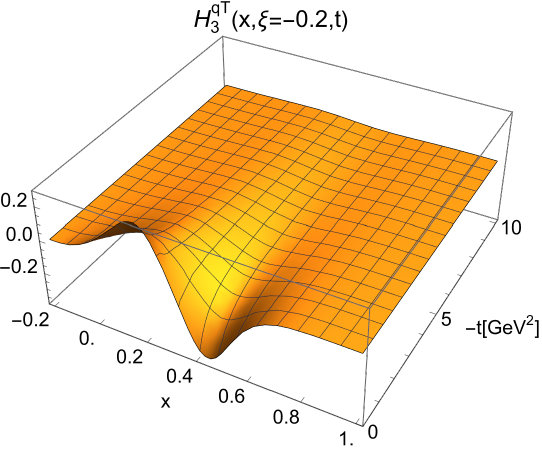}
	\includegraphics[width = 0.2 \textwidth]{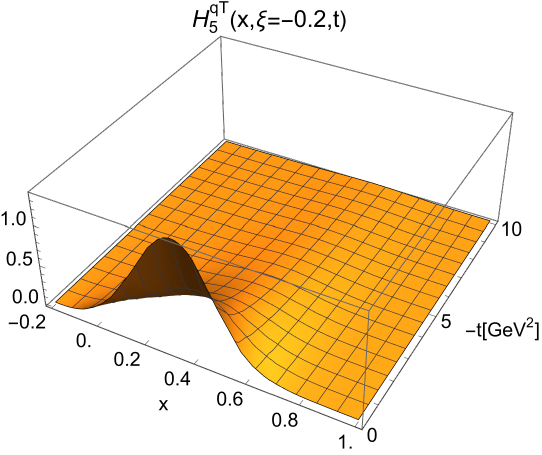}
	\includegraphics[width = 0.2 \textwidth]{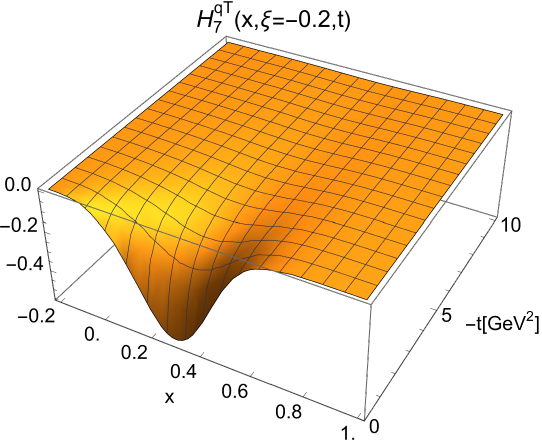}\\
	\includegraphics[width = 0.2 \textwidth]{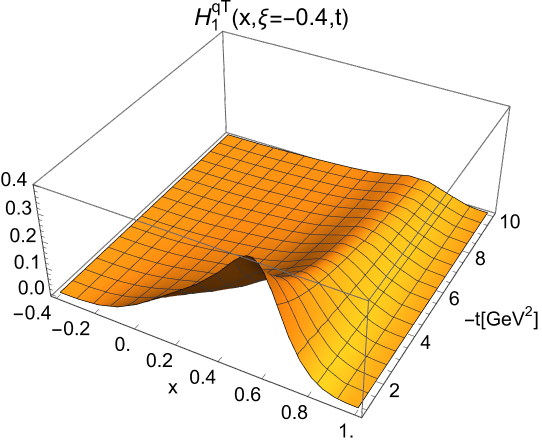}
	\includegraphics[width = 0.2 \textwidth]{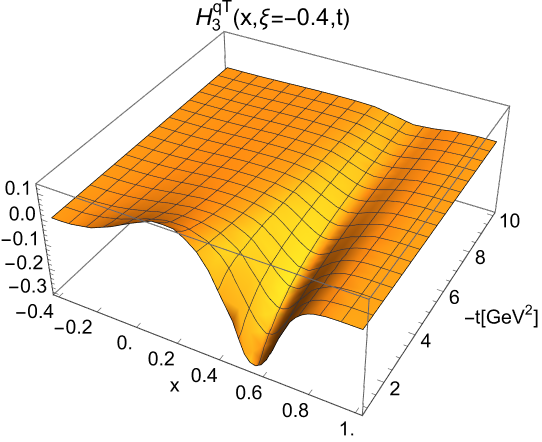}
	\includegraphics[width = 0.2 \textwidth]{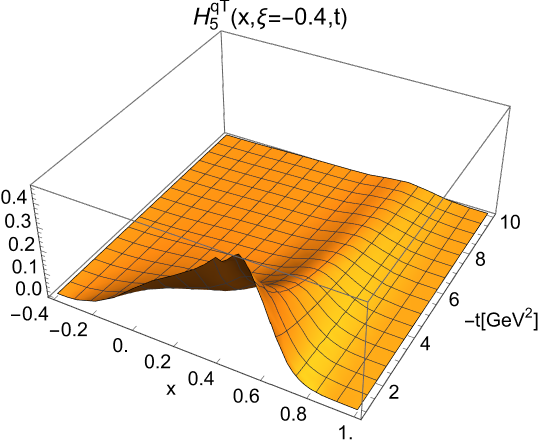}
	\includegraphics[width = 0.2 \textwidth]{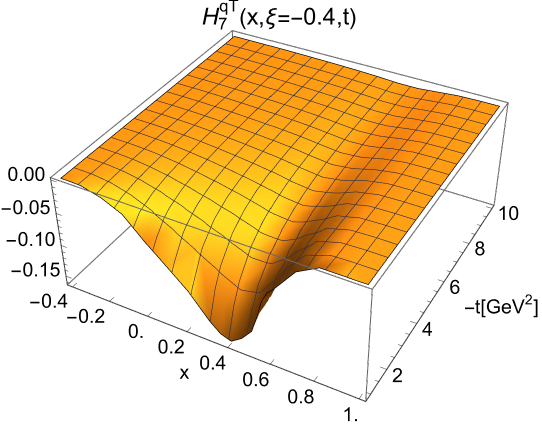}
	\caption{\small{The 3D $d$ quark transversity GPDs of $\Delta^+$ $H^{qT}_{1,3,5,7}$
	 at $\xi = 0, \, -0.2, \, -0.4$.}}
  \label{fig-tgpds}
\end{figure}
\noindent
the different quark GPDs 
in all the $\Delta$ isobars can be obtained just by counting
the corresponding quark number. 
With the approximation $c_2=c_3=0$, only $H^{qT}_{1,3,5,7}$, 
all of which respectively corresponds to the Lorentz structure of spin-1/2, 
survive and other transversity GPDs vanish as explained in Sec.~\ref{section-model}.
The non-zero transversity GPDs are shown in Fig.~\ref{fig-tgpds} 
as the functions of variables $x$ and $t$ with different skewness $\xi$, 
in which $\xi=(0,-0.2,-0.4)$, respectively. 
The constraint $|\xi| \le 1 /\sqrt{1-4M^2 /t}$ leads to a bound on 
the squared momentum transfer $-t \ge \frac{4 M^2 \xi^2}{1-\xi^2}$, from which one can get the corresponding $-|t|_\text{min} \sim (0, -0.2, -0.9)$.
Due to time reversal constraints in Eq.~\eqref{odd-even}, 
only the negative $\xi$ region is shown in the results.
We have verified the sum rules of the GPDs in Eq.~\eqref{quarksumrules}, 
$\int_{-1}^{1} \mathrm{d}x H^{qT}_3 (x,\xi,t)=0$.
Similar with unpolarized and longitudinal polarized GPDs~\cite{Fu:2023dea},
there is also a tendency that the maximums or minimums of $\xi$-even transversity GPDs $H^{qT}_{1,5,7}$ shift to large $x$ as $|\xi|$ increases.
In order to illustrate this character more intuitively and to observe the change 
with different $\xi$, the 2D cutting planes of $H_1^{qT}$ with different $\xi$ 
and the corresponding minimal $-t$ are listed in Fig.~\ref{fig-tgpd1}.
One can assume that the ratio of the parton momentum fraction equals 
the mass ratio, $i$.$e$. $\frac{x+|\xi|}{1-x} = \frac{m_q}{m_D}$, 
and this assumption determines a position 
$x_{\text{max}} = \frac{m_q + m_D |\xi|}{m_q + m_D}$.
The positions $x_{\text{max}} = 0.345 ,\, 0.476 ,\, 0.607$ respectively corresponds 
to $\xi=0 ,\, -0.2 ,\, -0.4$.
Furthermore, one can find that the $\xi$-even GPDs are concentrated 
near $x_{\text{max}}$. This character implies that the parton momenta 
are distributed by their masses and 
it agrees with our intuition.
Moreover, the GPDs corresponding to 
the unpolarized and longitudinaly-polarized PDFs
also have this feature~\cite{Fu:2023dea}.

In the forward limit, the transversity GPDs degenerate to transversity 
PDFs according to Eq.~\eqref{quarktransversitydistribution}. 
Due to our choice $c_2=c_3=0$, only the $h_1$ exists.
Moreover, the choice makes that only the spin-1/2 structure $i \sigma^{n i}g^{\alpha' \alpha}$ can be embodied in the transversity PDF and 
there is no contribution from $H^{qT}_2$ corresponding to the spin-1 structure $n^{[ \alpha' \nobracket} g^{\nobracket \alpha ] i}$.
Figure~\ref{fig-tpdf} shows the transversity PDF of $d$ quark in $\Delta^+$.

\begin{figure}[htbp]
	\centering
	\includegraphics[width=16cm]{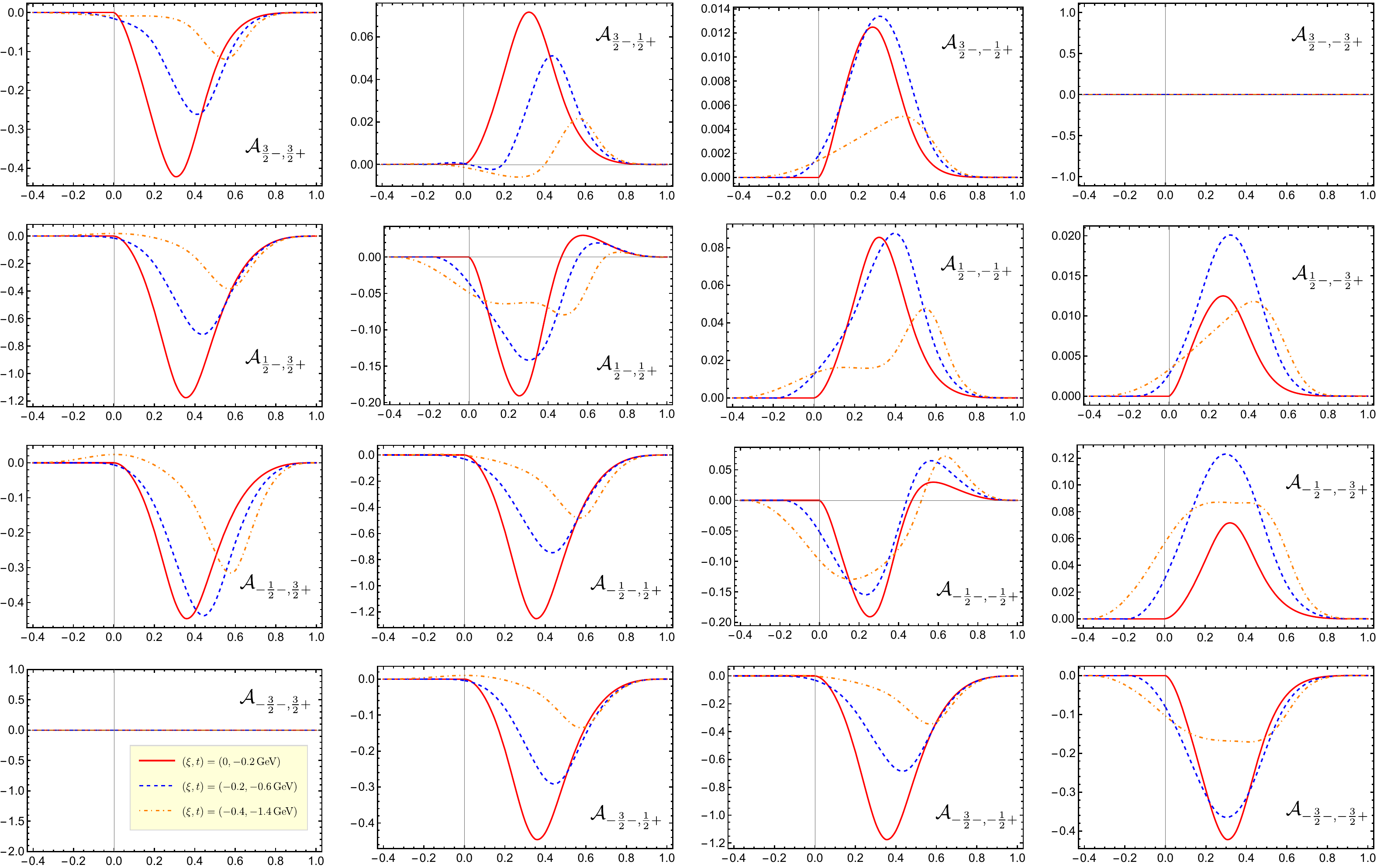}
	\caption{\small{The helicity-flip amplitudes of $d$ quark in $\Delta^+$, where the abscissa represents momentum fraction $x$.}}
	\label{fig-amp}
\end{figure}

As the combination of the GPDs, the quark helicity amplitude is another 
important quantity in the particle inner structure description. 
The quark amplitudes with helicity-flip can be defined as
\begin{equation}\label{helicityamplitude}
	\mathcal{A}^{q}_{\lambda' -, \lambda +}=\int \frac{\text{d} z^-}{2 \pi} e^{i x \left( P \cdot z \right)}
	\left.\left\langle p', \lambda' \left| \bar{\psi}\left( -\frac{1}{2}z \right) \frac{1}{4}\left( -i \sigma^{+1} + i i \sigma^{+2} \right)  \psi\left( \frac{1}{2}z \right)   \right|p, \lambda \right\rangle \right| _{z^+=0, \bm{z}_\perp=0},
\end{equation}
where the labels $-$ and $+$ are the emitted and re-absorbed quark 
helicities, respectively, and the explicit forms of the helicity-flip amplitudes are given in Ref.~\cite{Fu:2024kfx}.
According to the explicit forms in 
Ref.~\cite{Fu:2024kfx}, the helicity amplitudes can be expressed as $\mathcal{A}^{q}_{\lambda' -, \lambda +} = F(\zeta) \mathcal{A}'^q_{\lambda' -, \lambda +}$, where $F(\zeta)$ is the complex part and $\mathcal{A}'_{\lambda' -, \lambda +}$ is the real part.
The specific form of the complex part is $F(\zeta) = C^\zeta \theta(\zeta) + C^{*-\zeta} \theta(-\zeta)$ with $\zeta = \lambda' - \lambda +1 $, where $C^\zeta$ represents the $\zeta$ powers of $C$ and $\theta(\zeta)$ is the Heaviside step function with $\theta(0)=\frac{1}{2}$ and $F(0) = 1$.
The factor $F(\zeta)$ will be real when the helicity conserves, $\lambda' - \lambda = -1$ (i.e. the helicity conservation implies that the helicity transfer of the hadron corresponds to that of the parton).
Moreover, the factor $C$ carries all the complex phase information,
\begin{equation}
	\begin{split}
		C = \sqrt{\frac{1 - \xi}{1 + \xi}} \frac{| \bm{p}_{\bot} |}{M} e^{- i \phi} -
    \sqrt{\frac{1 + \xi}{1 - \xi}} \frac{| \bm{p}'_{\bot} |}{M} e^{- i \phi'}
    = - \frac{(\Delta+2\xi P)^x-i (\Delta+2\xi P)^y}{M \sqrt{1-\xi^2}},
	\end{split}
\end{equation}
where $| \bm{p}_{\bot} |e^{\pm i \phi} \equiv p^x \pm i p^y$ and $| \bm{p}'_{\bot} |e^{\pm i \phi'} \equiv p'^x \pm i p'^y$.
In the above equation, $C$ represents the transverse momentum 
transfer and also orbital angular momentum from the initial to 
the final hadronic states. One can obtain 
$|C|= \frac{\left|\bm{\Delta}_\perp\right|}{M \sqrt{1-\xi^2}}$ 
after removing the complex phase in the Breit frame.
Furtherly, the numerical results of the helicity-flip amplitudes without the complex phase are shown in Fig.~\ref{fig-amp}.
The bound $-t \ge \frac{4 M^2 \xi^2}{1-\xi^2}$ implies that $-t$ reaches its minimus when $\mathbf{\Delta}_\perp = \mathbf{0}$.
Consequently, under the condition that $-t = |t|_{min}$, the factor $C$ vanishes and the helicity amplitudes violating the helicity conservation will be zero.
Therefore, only the helicity amplitudes with helicity conservation are exist.
Moreover, to compare the helicity-flip amplitudes with different $\xi$, the specific variable values $(\xi, t) = (0, -0.2 \,\text{GeV}), \, (-0.2, -0.6 \,\text{GeV}), \, (-0.4, -1.4 \,\text{GeV})$ are selected in Fig.~\ref{fig-amp}.
Moreover, other helicity-flip amplitudes can be obtained by the constraints~\cite{Fu:2024kfx},
\begin{equation}
    \begin{split}
        \mathcal{A}^q_{-\lambda' - \mu', - \lambda - \mu}(P, \Delta, n) & = (-1)^{(\lambda'-\lambda)-(\mu'-\mu)} \mathcal{A}^{q *}_{\lambda' \mu', \lambda \mu}(P, \Delta, n), \\
        \mathcal{A}^q_{-\lambda' - \mu', - \lambda - \mu}(P, \Delta, n) & = (-1)^{(\lambda'-\lambda)-(\mu'-\mu)} \mathcal{A}^{q}_{\lambda \mu, \lambda' \mu'}(P, -\Delta, n),
    \end{split}
\end{equation}
from Hermiticity, parity and time reversal transformations.

\subsection{Tensor form factors}
\begin{figure}[h]
	\centering
	\includegraphics[height=4.5cm]{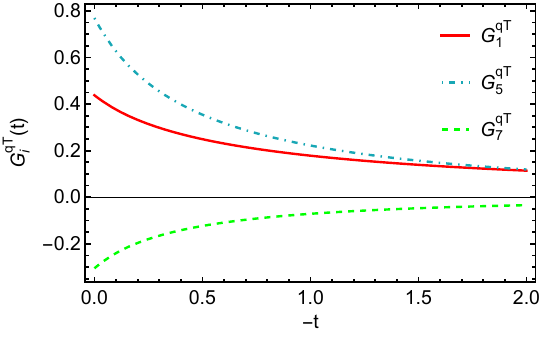}
	\caption{\small{The tensor form factors of $\Delta^+$ contributed by the $d$ quark.}}
	\label{fig-tffs}
\end{figure}

The sum rules in Eq.~\eqref{quarksumrules} derived from the Mellin 
moments of GPDs give the connections between the GPDs and the 
form factors.
Here, the matrix element of the local tensor current 
$\bar{\psi}(0) i \sigma^{\mu \nu} \psi(0)$ corresponding to the 
transversity GPDs can be decomposed in terms of the tensor form 
factors in Eq.~\eqref{tensorformfactors}.
According to the sum rules and the numerical GPDs, three non-zero 
tensor form factors are obtained and shown in Fig.~\ref{fig-tffs}.
Note that we just give the single quark contribution because of the 
unknown physical meaning of the local tensor current.
In particular, one can calculate the tensor charge $g_T$ from the integral of the transversity PDF $h_1(x)$~\cite{Gamberg:2021lgx} over the parton momentum fraction $x$:
\begin{equation}
    g_T = \delta u - \delta d,
\end{equation}
with
\begin{equation}
    \delta u = \int^1_0 d x \Big( h_1^u(x) - h_1^{\bar{u}}(x) \Big), \quad \delta d = \int^1_0 d x \Big( h_1^d(x) - h_1^{\bar{d}}(x) \Big),
\end{equation}
where $u$ and $d$ respectively represent up and down quarks.
There is no antiquark in our model $h_1^{\bar{u}}(x)=h_1^{\bar{d}}(x)=0$ and the spin and flavour wave functions are symmetric for the $\Delta$ resonance.
According to Eqs.~\eqref{quarktransversitydistribution} and \eqref{quarksumrules}, one therefore can obtain $\delta d = 2 \left[G^{qT}_1(0)-G^{qT}_2(0)\right]=0.876$ and $\delta u = 2 \delta d$ due to the number ratio between $u$ and $d$ quarks in hardon $\Delta^+$.
Therefore, we can obtain the tensor charge of the $\Delta$ resonance,
\begin{equation}
    g_T^{\Delta^{++}} = 2.628, \quad g_T^{\Delta^{+}} = 0.876, \quad g_T^{\Delta^{0}} = -0.876, \quad g_T^{\Delta^{-}} = -2.628,
\end{equation}
from the quark number, and the results can be predicted by other models and the lattice QCD.
Experimentally, one can extract the tensor charge using the transverse momentum dependent observables, like the Collins effect in semi-inclusive deep-inelastic scattering (SIDIS)~\cite{HERMES:2004mhh,JeffersonLabHallA:2011ayy} and semi-inclusive $e^+ e^- \to h \bar{h}$ (SIA, $h$ represents hadron here)~\cite{Belle:2008fdv,BESIII:2015fyw}, GPDs~\cite{Goldstein:2014aja}, and the hadron observables in SIDIS~\cite{HERMES:2008mcr,COMPASS:2012bfl,COMPASS:2014ysd} and SIA~\cite{Belle:2011cur,Cocuzza:2023oam}.

\section{\uppercase{Summary and Conclusion}}\label{section4}

Based on the diquark spectator model, we calculated the transversity
GPDs, the transversity PDF, the helicity-flip amplitudes and the tensor form factors of 
$\Delta^+$ for the first time.
According to our previous studies and our analyses here, the $c_{2,3}$ terms of the hadron-quark-diquark vertex mainly contribute to the high order physical quantities, like the magnetic-dipole and electric-quadrupole form factors.
In this work, the hadron-quark-diquark vertex is simplified
to be the $g^{\alpha \beta}$ form by neglecting the $c_{2,3}$ terms.
Then, only the GPDs corresponding to the Lorentz structures occurred in spin-1/2 transversity GPD definition are non-zero.
Meanwhile, our numerical results verify that this simplification is sufficient to describe the electric charge and axial charge form factors while magnetic-dipole form factor is acceptable.
Therefore, we believe that one can obtain the reasonable tensor charge form factor and the transversity PDF $h_1$. 

The non-zero transversity GPDs of $d$ quark in $\Delta^+$ are calculated and satisfy the corresponding sum rules.
The even terms $H^{qT}_{1,5,7}$ with respect to the skewness $\xi$ have the same feature, they mainly distributed around $x_{\text{max}} = \frac{m_q+m_D |\xi|}{m_q + m_D}$, with the unpolarized and longitudinal polarized ones.
In the forward limit, $\xi=0$ and $t=0$, the transversity PDF can be obtained from the transversity GPDs, and only the leading order $h_1$ exists.
Moreover, the helicity-flip amplitude as the combination of the GPDs is another significant physical quantity.
In the helicity amplitudes, the imaginary part is from the transverse momentum transfer, $i$.$e$. the orbital angular momentum from the initial to the final hadronic states.
The numerical results are calculated taking the absolute value of the transverse momentum transfer in the Breit frame.

Moreover, the Mellin moment connects the non-local transversity quark-quark operator $\bar{\psi}(-z /2) i \sigma^{n i} \psi(z /2)$ 
with the local tensor current $\bar{\psi}(0) i \sigma^{\mu \nu} \psi(0)$.
Meanwhile, the tensor form factors and the sum rules connecting the GPDs and the form factors have been derived in our previous work.
The numerical tensor form factors of $\Delta^+$ contributed by the $d$ quark are immediately obtained after integrating the parton momentum fraction $x$ and only $G^{qT}_{1,5,7}$ are non-zero.
Integral of $x$ in the transversity PDF $h_1$ and the tensor form factor in the forward limit give the tensor charge of the $d$ quark, $\delta_d = \int^1_0 d x \left( h_1^d(x) - h_1^{\bar{d}}(x) \right) = 2 \left[ G^{qT}_1(0)-G^{qT}_2(0) \right] = 0.876$ because there is no antiquark in our model.
Due to the symmetry of the spin and flavour wave functions for the $\Delta$ resonance, one can get the relation $\delta_u = 2 \delta_d$ in $\Delta^+$ and we can give the prediction of the tensor charge of $\Delta^+$, $g_T = 0.876$. Moreover, the tensor charge of other $\Delta$ resonances can be derived by the quark number.
And the tensor charge can be detected by various processes, such as SIDIS, SIA and the GPD processes.

\section{\uppercase{acknowledgments}}

This work is supported by the National Key R$\&$D Program of China under Contracts No.2023YFA1606703, and by the National Natural Science Foundation of China under Grants Nos. 12375142 and 12447121.
This work is also supported by the Gansu Province Postdoctor Foundation.

\bibliographystyle{unsrt}
\bibliography{refs}
\end{document}